# Why have no manifestations of the excitonic mechanism been detected in Ginzburg sandwiches?

Yu. A. Krotov

*P. N. Lebedev Institute of Physics, Russian Academy of Sciences, 117924 Moscow, Russia*

I. M. Suslov

*P. L. Kapitsa Institute of Physical Problems, Russian Academy of Sciences, 117334 Moscow, Russia*

Using spatially inhomogeneous Éliashberg equations in the local-interaction limit, an exact solution of the problem of the superconducting transition temperature in a Ginzburg sandwich (a superconducting film coated with a dielectric layer containing Bose-type excitations, i.e., excitons) in the first order in $a/L$ (where $a$ is the interatomic distance and $L$ is the film thickness) has been obtained. The result has been found to be independent of the exciton frequency. The excitonic mechanism appears only in second order in $a/L$ since both components of the Cooper pair should enter a layer of thickness $\sim a$ in order to interact through the exchange of excitons. Numerical estimates indicate that manifestations of the excitonic mechanism are practically undetectable in systems with $L \gg a$. Calculations for the model with a narrow-gap and a wide-gap dielectric have been performed and compared to experimental data. © *1997 American Institute of Physics.* [S1063-7761(97)02402-5]

## 1. INTRODUCTION

In 1964 in his famous paper,[1] Ginzburg set forth a new method for creating high-temperature superconductors. If a thin metal film is coated with a layer of a dielectric (Fig. 1) containing high-frequency boson excitations, i.e., excitons, whose frequency $\omega_{\mathrm{ex}}$ is considerably higher than the phonon frequency $\omega_{\mathrm{ph}}$ in the metal, the combination of the finite electron density of states on the interface and the high excitation frequency should lead, according to the BCS formula, to a high local value of the superconducting transition temperature $T_c$. The theory of Ginzburg sandwiches has been developed by many authors (see Ch. 8 in Ref. 2 and references therein), but the available estimates of $T_c$ are unsatisfactory because all these theories ignore the problems related to the spatial inhomogeneity of sandwiches. All of them were based on the use of the BCS or MacMillan-type formulas and rough estimates of their parameters. Below we shall demonstrate that such an approach leads to qualitatively erroneous results.

Following the terminology of Ref. 2, we define sandwiches as structures manufactured using the appropriate technology, such that their metal film thickness $L$ is essentially larger than the interatomic distance $a$.[1)] Structures with $L \sim a$ should be treated as quasi-two-dimensional, and this topic is beyond the scope of this paper. Besides, we assume that the superconductivity in the film is three-dimensional since the predicted surface superconductivity of Tamm states[1,5] has not been detected with certainty in any material.

It is clear from general principles that the difference between $T_c$ in a sandwich and $T_{c0}$ in the bulk material of the film should be proportional to $a/L$:

$$\frac{\delta T_c}{T_{c0}} \equiv \frac{T_c - T_{c0}}{T_{c0}} = C \frac{a}{L}. \qquad (1)$$

All the existing theories[2] are based on the assumption that, in the formal limit $\omega_{\mathrm{ex}} \to \infty$, the factor $C$ should diverge, and its large value should compensate for the smallness of the ratio $a/L$ or, at least, make the exciton-mediated interaction dominant over all other effects, which yield $C \sim 1$. In this paper, however, we demonstrate that

$$C(\omega_{\mathrm{ex}}) = \mathrm{const} \quad \text{for} \quad \omega_{\mathrm{ex}} \gtrsim \omega_{\mathrm{ph}}. \qquad (2)$$

This result, however strange it may seem at first sight, is natural. In order to interact through the exchange of excitons, both components of a Cooper pair must reach a layer with thickness of the order of $a$, and the probability of this event is $\sim (a/L)^2$, hence the excitonic mechanism should not appear to first order in $a/L$. If the term quadratic in $a/L$ is considered, after the dimensionless interaction constant $\lambda_0$ in the bulk metal is factored out, we have

$$\frac{\delta T_c}{T_{c0}} = \frac{A}{\lambda_0} \frac{a}{L} + \frac{B(\omega_{\mathrm{ex}})}{\lambda_0} \left(\frac{a}{L}\right)^2 + \ldots, \qquad (3)$$

and the coefficients in this formula can be estimated as[2)]

$$B(\omega_{\mathrm{ex}}) = B_0 + B_1 \lambda_0 \ln \frac{\omega_{\mathrm{ex}}}{\omega_{\mathrm{ph}}}, \quad A, B_0, B_1 \sim 1, \qquad (4)$$

i.e., the coefficient of $(a/L)^2$ in fact diverges as $\omega_{\mathrm{ex}} \to \infty$. The factors $\lambda_0^{-1}$ arise in Eq. (3) because the variation of the BCS formula $T_c \sim \bar{\omega} \exp(-1/\lambda)$ with respect to $\bar{\omega}$ and $\lambda$ yields $\delta T_c/T_{c0}$ proportional to $\delta \bar{\omega}/\bar{\omega}$ and $\delta\lambda/\lambda_0^2$, respectively, i.e., the relative change in $\lambda$ is multiplied by the factor $\lambda_0^{-1}$, as compared to the relative change in $\bar{\omega}$. According to Eqs. (3) and (4), the ratio of the contribution of the excitonic mechanism to the total change in $T_c$ is

$$\frac{(\delta T_c)_{\mathrm{ex}}}{(\delta T_c)_{\mathrm{tot}}} \sim \frac{a}{L} \lambda_0 \ln \frac{\omega_{\mathrm{ex}}}{\omega_{\mathrm{ph}}}. \qquad (5)$$

## 2. SPATIALLY INHOMOGENEOUS ELIASHBERG EQUATIONS

Consider the Hamiltonian of electron–phonon interaction in the form

$$H_{int} = -\int d\mathbf{r}\hat{\psi}_\sigma^+(\mathbf{r})\mathbf{u}_n\mathbf{g}_n(\mathbf{r})\hat{\psi}_\sigma(\mathbf{r}), \tag{7}$$

where $\hat{\psi}_\sigma^+$ and $\hat{\psi}_\sigma$ are electron operators, $\mathbf{u}_n$ is the displacement vector of the $n$th ion, and $\mathbf{g}_n(\mathbf{r})$ is the deformation potential which in the rigid-ion approximation takes the form[12]

$$\mathbf{g}_n(\mathbf{r}) = \nabla U_n(\mathbf{r} - \mathbf{R}_n), \tag{8}$$

where $U_n(\mathbf{r})$ is the potential of the $n$th ion and $\mathbf{R}_n$ is its equilibrium position. Following the standard procedure,[10] we obtain the spatially inhomogeneous Éliashberg equations $[x=(\mathbf{r},\tau)]$

$$\left(-\frac{\partial}{\partial\tau} - \hat{H}_0 + \mu\right)G(x,x') = \delta(x-x') - \int dx_1 G(x,x_1)$$
$$\times D(x,x_1)G(x,x') + \int dx_1$$
$$\times F(x,x_1)D(x,x_1)F^+(x_1,x'),$$

$$\left(-\frac{\partial}{\partial\tau} - \hat{H}_0 + \mu\right)F(x,x') = -\int dx_1 G(x,x_1)D(x,x_1)$$
$$\times F(x_1,x') - \int dx_1 F(x,x_1)$$
$$\times D(x,x_1)G(x',x_1), \tag{9}$$

where $G$ and $F$ are the normal and anomalous Green's function and $\mu$ is the chemical potential. Unlike Eq. (35.2) in Ref. 10, all the functions in Eq. (9) depend on the two coordinates, not just on their difference, the operator $\hat{p}^2/2m$ is replaced by a one-particle Hamiltonian $\hat{H}_0$ of general form, and the coupling constant $g$ is included in the definition of the phonon Green's function:

$$D(x,x') = \sum_{\alpha,\alpha'}\sum_{n,n'} g_n^\alpha(\mathbf{r})g_{n'}^{\alpha'}(\mathbf{r}')D_{nn'}^{\alpha\alpha'}(\tau-\tau'), \tag{10}$$

where $D_{nn'}^{\alpha\alpha'}$ is the Green's function in the site representation, which can be expressed in terms of eigenvectors $B_\alpha^{(s)}(n)$ and eigenvalues $\omega_s^2$ of the dynamic operator matrix:[13]

$$D_{nn'}^{\alpha\alpha'}(\Omega) = -\frac{\hbar}{\sqrt{M_nM_{n'}}}\sum_s \frac{B_\alpha^{(s)}(n)B_{\alpha'}^{(s)}(n')}{\Omega^2+\omega_s^2}, \tag{11}$$

$M_n$ is the mass of the $n$th ion, and $\Omega$ is the Matsubara frequency.

In order to determine $T_c$, we must linearize Eq. (9) in $F$. If we rewrite these equations in the symbolic form

$$\left(-\frac{\partial}{\partial\tau} - \hat{H}_0 + \mu + GD\right)G = 1,$$

$$\left(-\frac{\partial}{\partial\tau} - \hat{H}_0 + \mu + GD\right)F = -FDG, \tag{12}$$

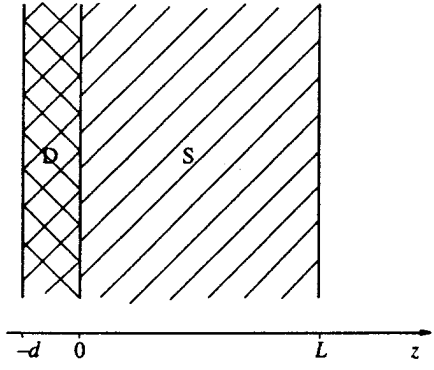

FIG. 1. A Ginzburg sandwich is a thin superconducting film S with a deposited dielectric layer D with high-frequency bosons excitations, whose exchange should increase $T_c$ considerably.

The limitations due to the Tolmachev logarithm[2] lead to the inequality $\omega_{ex}/\omega_{ph} \lesssim 10^2$, and for typical values $\lambda_0 = 0.2$–$0.3^{3)}$ we have $\lambda_0 \ln(\omega_{ex}/\omega_{ph}) \sim 1$, hence the excitonic contribution is always small at $L \gg a$. This means that any attempts to detect the excitonic effects in the "small correction regime" are doomed to failure: when a metal film is coated with a dielectric, the change in $T_c$ may be controlled by all effects, except the exchange of excitons. This is, apparently, the main reason why this effect has not been detected in experiments.

A formula for $T_c$, naturally, cannot be derived for an arbitrary spatially inhomogeneous system, but this is possible for the case of a local spatial inhomogeneity, when its dimension $d$ is smaller than the coherence length $\xi_0$ or the total system dimension $L$ (if $L \lesssim \xi_0$). These formulas were derived earlier[6,7] and used to study the localization of the order parameter localization, quantum oscillations of $T_c$, the contribution of an interface between two materials to $T_c$ as a function of material parameters, etc.[6–9] All these studies used the Gor'kov equation,[10,11] which does not allow for the spatial dependence of the cut-off frequency $\bar{\omega}$. In the weak-coupling regime, this dependence expressed by $\bar{\omega}(\mathbf{r})$ usually leads only to small corrections determined by the parameter

$$\lambda_0 \ln\frac{\bar{\omega}_{max}}{\bar{\omega}_{min}} \ll 1. \tag{6}$$

In the case of a large disparity in the frequencies, $\bar{\omega}_{max} \gg \bar{\omega}_{min}$, the condition (6) may be violated even for $\lambda_0 \ll 1$, and this is the case in a Ginzburg sandwich.

In the present study, formulas for $T_c$ similar to those in Refs. 6 and 7 are derived from the spatially inhomogeneous Éliashberg equations.[10] Since Ginzburg's concept does not depend on the nature of the high-frequency bosons, we have used the Éliashberg equations for the case of electron–phonon interaction. Their structure is, in fact, identical for any bosons with frequencies small in comparison to the Fermi energy $\epsilon_F$. This statement especially applies to the limit of local interaction (Sec. 3), in which no specific information about phonons is essential.

we note that $G$ is the Green's function for the operator in parentheses and can rewrite the second equation in (12) as $F = -GFDG$. In the explicit form, after transformation to the Matsubara representation and complex conjugation, we have

$$F_\omega^+(\mathbf{r},\mathbf{r}') = -T\sum_\Omega \int d\mathbf{r}_1 \int d\mathbf{r}_2 G_{-\omega}(\mathbf{r}_1,\mathbf{r})$$
$$\times D_\Omega(\mathbf{r}_1,\mathbf{r}_2) F_{\omega-\Omega}^+(\mathbf{r}_1,\mathbf{r}_2) G_\omega(\mathbf{r}_2,\mathbf{r}'). \quad (13)$$

Let us introduce the order parameter

$$\Delta_\omega(\mathbf{r},\mathbf{r}') = -T\sum_\Omega F_{\omega-\Omega}^+(\mathbf{r},\mathbf{r}') D_\Omega(\mathbf{r},\mathbf{r}'), \quad (14)$$

and rewrite Eq. (13) in the form

$$\Delta_\omega(\mathbf{r},\mathbf{r}') = -T\sum_{\omega'} \int d\mathbf{r}_1 d\mathbf{r}_2 D_{\omega-\omega'}(\mathbf{r},\mathbf{r}')$$
$$\times G_{-\omega'}(\mathbf{r}_1,\mathbf{r}) G_{\omega'}(\mathbf{r}_2,\mathbf{r}') \Delta_{\omega'}(\mathbf{r}_1,\mathbf{r}_2). \quad (15)$$

In Eqs. (13)–(15) $\Omega$ is the boson frequency, and $\omega$ and $\omega'$ are fermion frequencies. Note that Eq. (15) contains only renormalized Green's functions.

## 3. THE LOCAL-INTERACTION LIMIT

Equation (15) has a form similar to that of the Gor'kov equation[10,11] and reduces to the latter if two approximations typical of the BCS theory are used:

$$D_{\omega-\omega'}(\mathbf{r},\mathbf{r}') \rightarrow -V_{\omega-\omega'}(\mathbf{r}) \delta(\mathbf{r}-\mathbf{r}'), \quad (16)$$

$$V_{\omega-\omega'}(\mathbf{r}) \rightarrow V(\mathbf{r}) \theta(\bar\omega-|\omega|) \theta(\bar\omega-|\omega'|) \quad (17)$$

[as a result, $\Delta_\omega(\mathbf{r},\mathbf{r}') \rightarrow \Delta(\mathbf{r}) \delta(\mathbf{r}-\mathbf{r}') \theta(\bar\omega-|\omega|)$]. Equation (17) means that the spatial dependence of the cut-off frequency is ignored, and it will not be used further. This does not cause any complications because all the relevant equations can be solved by removing the logarithmic singularity (Ref. 2, p. 90).

The approximation expressed by Eq. (16) corresponds to the physically transparent local-interaction limit and has several advantages: (a) it yields simple and easily understandable results; (b) it does not demand a specification of the Fermi surface shape; (c) it does not demand detailed information about the electron–phonon interaction since, in fact, an interaction constant $V_\omega(\mathbf{r})$ which is an arbitrary function of the frequency and coordinates is introduced into Eq. (16), and so the generalization to other types of interaction is possible; (d) the structure of the expression for $T_c$ is identical to that derived from the Gor'kov equation, and earlier results[6–9] can be automatically generalized to the case of the cut-off frequency depending on coordinates. The absence of the effect of the excitonic mechanism to lowest order in $a/L$ can also be proved with due account of the nonlocality, but the expressions in this case would be too lengthy.

We should stress that the local interaction limit is a physical approximation and cannot be introduced by a mathematically rigorous procedure. In fact, if the function $D_\omega(\mathbf{r},\mathbf{r}')$ is assumed to be short-range and expressed as

$$D_\omega(\mathbf{r},\mathbf{r}') \approx \delta(\mathbf{r}-\mathbf{r}') \int d\mathbf{r}' D_\omega(\mathbf{r},\mathbf{r}'), \quad (18)$$

the integral on the right-hand side is zero because the integral of $\mathbf{g}_n(\mathbf{r})$ vanishes since the deformation potential is generated by redistribution of charges and can be described as a superposition of fields generated by dipoles. In the rigid-ion approximation it follows directly from Eq. (8). The local approximation is resonable from the physical viewpoint because the expression for $T_c$ is, in effect, determined by the integral in Eq. (18) over the region $|\mathbf{r}'| \lesssim k_F^{-1}$, where $k_F$ is the Fermi momentum. This can be proved by taking the result for the spatially homogeneous case in Ref. 2, Ch. 4.

If we assume the approximation described by Eq. (16), we have $\Delta_\omega(\mathbf{r},\mathbf{r}') = \Delta_\omega(\mathbf{r}) \delta(\mathbf{r}-\mathbf{r}')$, and Eq. (15) takes the form

$$\Delta_\omega(\mathbf{r}) = T\sum_{\omega'} V_{\omega-\omega'}(\mathbf{r}) \int d\mathbf{r}' K_{\omega'}(\mathbf{r},\mathbf{r}') \Delta_{\omega'}(\mathbf{r}'), \quad (19)$$

where

$$K_\omega(\mathbf{r},\mathbf{r}') = G_{-\omega}(\mathbf{r}',\mathbf{r}) G_\omega(\mathbf{r}',\mathbf{r}). \quad (20)$$

If the system is invariant under time reversal, the kernel $K_\omega(\mathbf{r},\mathbf{r}')$ is symmetric with respect to the exchange of $\mathbf{r}$ and $\mathbf{r}'$ and is positive. If for $G_\omega(\mathbf{r},\mathbf{r}')$ one-particle Green's functions are used, the following sum rule applies to the kernel:[11]

$$\int d\mathbf{r}' K_\omega(\mathbf{r},\mathbf{r}') = \frac{\pi}{|\omega|} N(\mathbf{r}), \quad (21)$$

where $N(\mathbf{r})$ is the local density of states at the Fermi level,

$$N(\mathbf{r}) = \sum_n |\psi_n(\mathbf{r})|^2 \delta(\epsilon_F - \epsilon_n), \quad (22)$$

determined by the one-particle eigenfunctions $\psi_n(\mathbf{r})$ and eigenvalues $\epsilon_n$. With due account of interaction effects, Eq. (21) can be considered as a definition of the local density of states $N(\mathbf{r})$. In the spatially homogeneous case this parameter (independent of $\mathbf{r}$) enters in the BCS formula.

## 4. THE EXPRESSION FOR $T_c$ IN THE CASE OF LOCAL SPATIAL INHOMOGENEITY

Suppose that the system varies as a function of $z$, and the inhomogeneity is localized in the region $|z| \lesssim d$. Since $\Delta_\omega(\mathbf{r})$ is independent of the longitudinal coordinate $\mathbf{r}_\parallel$, Eq. (19) takes the form

$$\boldsymbol{\Delta}(z) = \int dz' \hat{Q}(z,z') \boldsymbol{\Delta}(z'), \quad (23)$$

where $\boldsymbol{\Delta} = (\Delta_{\omega_1}, \Delta_{\omega_2}, \ldots)$. If the system transverse dimension satisfies $L \ll \xi_0$, the solution can be sought in the form[6,7]

$$\boldsymbol{\Delta}(z) = \boldsymbol{\psi} + \boldsymbol{\Delta}_0(z), \quad (24)$$

where the function $\boldsymbol{\psi}$ is independent of $z$ and $\boldsymbol{\Delta}_0(z)$ is localized in the region $|z| \lesssim d$. Substituting Eq. (24) into (23), we obtain

$$\boldsymbol{\psi} = \int dz' \hat{Q}(\infty,z') \boldsymbol{\psi} + \int dz' \hat{Q}(\infty,z') \boldsymbol{\Delta}_0(z'), \quad (25)$$

$$\Delta_0(z) = \int dz' [\hat{Q}(z,z') - \hat{Q}(\infty,z')] \psi + \int dz' [\hat{Q}(z,z') - \hat{Q}(\infty,z')] \Delta_0(z'). \quad (26)$$

In deriving these equations we have taken into account that for $|z| \gtrsim d$ the kernel $\hat{Q}(z,z')$ is independent of $z$ and equals $\hat{Q}(\infty,z')$. The sum rule (21) implies the estimate $\hat{Q} \sim 1/L$, and the second terms on the right-hand side of Eqs. (25) and (26) are small $\sim d/L$. In order to calculate $T_c$ with to $\sim d/L$ inclusive, we can omit the second term on the right-hand side of Eq. (26) and substitute the resulting $\Delta_0(z)$ into Eq. (25). Given that $K_\omega(\infty,z') \approx L^{-1} \int dz K_\omega(z,z')$, we obtain with due account of Eq. (21) an equation for $\psi$ in the explicit form

$$\psi_\omega = \pi T \sum_{\omega'} \frac{L(\omega,\omega')}{|\omega'|} \psi_{\omega'}, \quad (27)$$

$$L(\omega,\omega') = V_{\omega-\omega'}(\infty) N(\infty)$$
$$+ \frac{1}{L} \int dz \pi T \sum_{\omega''} \frac{1}{|\omega''|} V_{\omega-\omega''}(\infty) N(z)$$
$$\times [V_{\omega''-\omega'}(z) N(z) - V_{\omega''-\omega'}(\infty) N(\infty)]. \quad (28)$$

Equation (27) can be solved by removing the logarithmic singularity.[4] By taking advantage of the fact that the summation over Fermi frequencies yields

$$\pi T \sum_{|\omega|<\bar{\omega}} \frac{1}{|\omega|} = \ln \frac{1.14\bar{\omega}}{T}, \quad (29)$$

we transform Eq. (27) to

$$\psi_\omega = L(\omega,\omega_0) \psi_{\omega_0} \ln \frac{1.14\bar{\omega}}{T} + f(\omega), \quad (30)$$

where $\omega_0 = \pi T$, and the function

$$f(\omega) = \pi T \sum_{|\omega'|>\bar{\omega}} \frac{L(\omega,\omega') \psi_{\omega'}}{|\omega'|}$$
$$+ \pi T \sum_{|\omega'|<\bar{\omega}} \frac{L(\omega,\omega') \psi_{\omega'} - L(\omega,\omega_0) \psi_{\omega_0}}{|\omega'|} \quad (31)$$

is introduced. After setting $\omega = \omega_0$ and $L(\omega_0,\omega_0) \approx L(0,0)$ in Eq. (30), we have the expression for $T_c$:

$$T_c = 1.14 \bar{\omega} e^{-1/L(0,0)}, \quad (32)$$

where $\bar{\omega}$ is defined by the condition $f(\omega_0) = 0$. After replacing summation by integration in Eq. (31) and substituting $\psi_\omega$ in the lowest-order approximation [i.e., neglecting $f(\omega)$ in Eq. (30)], we obtain for $\bar{\omega}$

$$\ln \bar{\omega} = -\frac{1}{L(0,0)^2} \int_0^\infty \ln \omega [L(0,\omega) L(\omega,0)]'_\omega d\omega. \quad (33)$$

Substitution of Eq. (28) into (32) and (33) and expansion in terms of $d/L$ yield the variation in $T_c$ relative to $T_{c0}$ in the spatially homogeneous system

$$\frac{\delta T_c}{T_{c0}} = \frac{1}{\lambda_0^3 L} \int dz W_0 N(z) [W(z) N(z) - W_0 N_0]; \quad (34)$$

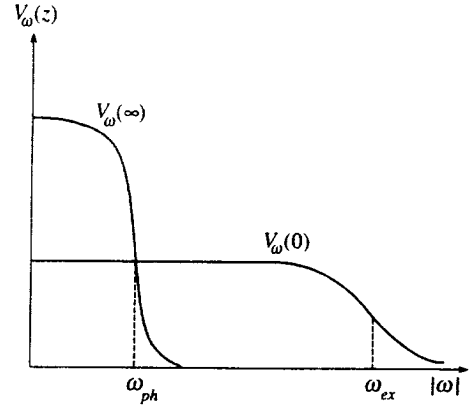

FIG. 2. Typical behavior of the parameter $V_\omega(z)$ as a functions of $\omega$ deep inside the metal film ($z=\infty$) and on the metal–dielectric interface ($z=0$).

here $\lambda_0 = V_0(\infty) N(\infty)$, $W_0 = W(\infty) = V_0(\infty)$, and the function

$$W(z) = V_0(z) - 2 N(\infty) \int_0^\infty d\omega \ln \frac{\omega}{\bar{\omega}} [V_\omega(\infty) V_\omega(z)]'_\omega, \quad (35)$$

is introduced, where $\bar{\omega}$ is calculated in the zeroth approximation,

$$\ln \bar{\omega} = -\frac{1}{V_0(\infty)^2} \int_0^\infty d\omega \ln \omega [V_\omega(\infty)^2]'_\omega, \quad (36)$$

corresponding to a spatially homogeneous system [so that $T_{c0} = 1.14\bar{\omega} \exp(-1/\lambda_0)$]. If $V_\omega(z)$ is a step function of the frequency,

$$V_\omega(z) = V(z) \theta(\bar{\omega}(z) - |\omega|), \quad (37)$$

it follows from Eq. (36) that $\bar{\omega} = \bar{\omega}(\infty)$, and Eq. (35) yields

$$W(z) = V(z) \left[ 1 + \lambda_0 \ln \frac{\min\{\bar{\omega}(\infty), \bar{\omega}(z)\}}{\bar{\omega}(\infty)} \right]. \quad (38)$$

For $\bar{\omega}(z) = $ const we have $W(z) = V(z)$, and Eq. (34) becomes identical to the result obtained in Ref. 7. It becomes clear that the spatial dependence of the cut-off frequency does not change the structure of Eq. (34), but only replaces $V(z)$ with a more complicated function $W(z)$.

For a Ginzburg sandwich, the main contribution to the integral in Eq. (34) comes from the region $|z| \lesssim a$ near the interface, and the relative variation of $T_c$ is $\sim a/L$. It is essential that the function $V_\omega(z)$, which contains information about the exciton frequency $\omega_{ex}$ at $|z| \lesssim a$ (Fig. 2), is multiplied in Eq. (35) by the function $V_\omega(\infty)$, which decreases fast at $|\omega| \gtrsim \omega_{ph}$. As a result, $\omega_{ex}$ is not included in Eqs. (34) and (35), which determine $T_c$. In the approximation described by Eq. (37), this directly follows from Eq. (38). This approximate result holds for all orders in $\lambda_0$ [Eqs. (34) and (35) were derived via iterations in this parameter]. Specifically, consider the eigenvalue equation

$$\nu \psi_\omega = \pi T \sum_{\omega'} \frac{V_{\omega-\omega'}(\infty) N(\infty)}{|\omega'|} \psi_{\omega'}, \quad (39)$$

which is identical to Eq. (27) for a spatially homogeneous system at $\nu=1$. If $\nu(T)$ is the maximal eigenvalue of Eq. (39), $T_{c0}$ is determined by the condition $\nu(T_{c0})=1$. Let $\bar{\psi}_\omega$ and $\bar{\bar{\psi}}_\omega$ be the solution to Eq. (39) and its adjoint solution at $\nu=1$. Then the perturbation calculation in the parameter $\sim d/L$ in Eq. (28) yields

$$\delta T_c = -\frac{1}{\nu'(T_{c0})\lambda_0^2 L}\int dz W_0 N(z)[W(z)N(z)-W_0 N_0] \quad (40)$$

with the function $W(z)$ defined by the equation

$$W(z) = \frac{\lambda_0}{\pi T \Sigma_\omega |\omega|^{-1}(\bar{\psi}_\omega)^2} \pi T$$

$$\times \sum_{\omega'} \frac{\bar{\psi}_{\omega'}}{|\omega'|} \pi T \sum_{\omega''} \frac{\bar{\psi}_{\omega''}}{|\omega''|} V_{\omega''-\omega'}(z), \quad (41)$$

using $\bar{\bar{\psi}}_\omega = |\omega|^{-1}\bar{\psi}_\omega$. It follows from the analysis of Eq. (39) that $\bar{\psi}_\omega$ decreases rapidly as a function of $\omega$ beyond $\omega_{ph}$, therefore the summation over the frequency in Eq. (41) is limited to the region $|\omega'|\lesssim\omega_{ph}$, $|\omega''|\lesssim\omega_{ph}$, and the frequency $\omega_{ex}$ in the function $V_{\omega''-\omega'}(z)$ does not affect the final result.

## 5. ESTIMATE OF THE GINZBURG EFFECT

By iterating Eqs. (25) and (26) up to second order in $d/L$,

$$\psi = \int dz'\hat{Q}(\infty,z')\psi + \int dz'\int dz''\hat{Q}(\infty,z')[\hat{Q}(z',z'')$$

$$-\hat{Q}(\infty,z'')]\psi + \int dz'\int dz''\int dz'''\hat{Q}(\infty,z')$$

$$\times[\hat{Q}(z',z'')-\hat{Q}(\infty,z'')][\hat{Q}(z'',z''')-\hat{Q}(\infty,z''')]\psi, \quad (42)$$

we obtain Eq. (27) with a function $L(\omega,\omega')$ differing from that defined by Eq. (28) by an additional term $\sim(d/L)^2$, which leads to a second-order correction to $T_c$:

$$\left(\frac{\delta T_c}{T_{c0}}\right)_2 = \frac{1}{\lambda_0}\frac{1}{V_0(\infty)}\frac{1}{L}\int dz'dz''N(z')\pi T$$

$$\times \sum_{\omega'}\frac{V_{\omega'}(\infty)}{|\omega'|} T\sum_{\omega''}[V_{\omega'-\omega''}(z')K_{\omega''}(z',z'')$$

$$-V_{\omega'-\omega''}(\infty)K_{\omega''}(\infty,z'')]\pi T\sum_{\omega'''}\frac{V_{\omega'''}(\infty)}{|\omega'''|}$$

$$\times[V_{\omega''-\omega'''}(z'')N(z'')-V_{\omega''-\omega'''}(\infty)N(\infty)]. \quad (43)$$

The summation over $\omega'$ and $\omega'''$ is limited to the region $|\omega'|,|\omega'''|\lesssim\bar{\omega}\sim\omega_{ph}$. By performing the summation with logarithmic accuracy and separating the contribution of the high-frequency region, we obtain the change in $T_c$ due to the exciton-mediated interaction:

$$\left(\frac{\delta T_c}{T_{c0}}\right)_{ex} = \frac{1}{\lambda_0^3 L}V_0(\infty)\int dz'dz''N(z')N(z'')T$$

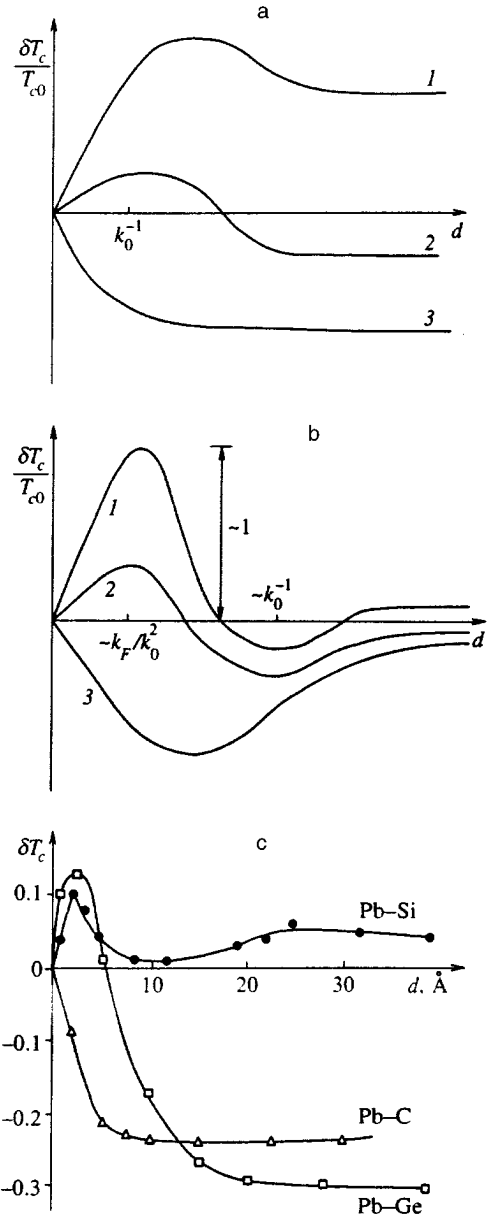

FIG. 3. (a) Calculated $T_c$ in a metal–dielectric sandwich as a function of the dielectric thickness for a narrow-gap dielectric ($U-\epsilon_F\ll\epsilon_F$); curves 1, 2, and 3 correspond to $W_1/W_0>7/2$, $3/2<W_1/W_0<7/2$, and $W_1/W_0<3/2$; (b) similar curves for a wide-gap dielectric ($U\gg\epsilon_F$) in the cases (curve 1) $W_1/W_0>3k_0^2/k_F^2$, (2) $3k_0^2/k_F^2>W_1/W_0>3k_0^2/k_F^2$, and (3) $W_1/W_0<k_0^2/2k_F^2$; (c) experimental curves of $T_c$ as a function of the dielectric thickness $d$ in Pb–Si, Pb–Ge, and Pb–C sandwiches.[14]

$$\times\sum_{|\omega|>\bar{\omega}}V_\omega(z')K_\omega(z',z'')V_\omega(z''). \quad (44)$$

If the local density of states $N(z)$ in the sandwich (Fig. 1) varies near the interface faster than $V_\omega(z)$, then, using the sum rule, we can assume that

$$K_\omega(z,z') \simeq \frac{\pi N_0}{|\omega|L}\theta(z)\theta(z'), \quad N(z)=N_0\theta(z); \quad (45)$$

hence

$$\left(\frac{\delta T_c}{T_{c0}}\right)_{\text{ex}} \simeq \frac{1}{\lambda_0^2} \frac{N_0^2}{L^2} \int_0^L dz' \int_0^L dz'' \pi T \sum_{|\omega|>\bar{\omega}} \frac{V_\omega(z')V_\omega(z'')}{|\omega|}. \quad (46)$$

Since the integrand is a logarithmic function of the cut-off frequency, Eq. (46) is valid in a fairly large region. In the simplest case, when

$$V_\omega(z) = \begin{cases} V_0 \theta(\omega_{\text{ph}} - |\omega|), & z > a, \\ V_1 \theta(\omega_{\text{ex}} - |\omega|), & z < a, \end{cases} \quad (47)$$

we have

$$\left(\frac{\delta T_c}{T_{c0}}\right)_{\text{ex}} \simeq \left(\frac{a}{L}\right)^2 \frac{V_1^2}{V_0^2} \ln \frac{\omega_{\text{ex}}}{\omega_{\text{ph}}}. \quad (48)$$

Since it is unlikely that the coupling constant $V_1$ for high-frequency excitations is larger than for low-frequency excitations ($V_0$), the contribution of the excitonic mechanism given by Eq. (48) is always smaller than the main contribution $\sim a/\lambda_0 L$ determined by Eq. (34).

## 6. MODEL CALCULATIONS TO FIRST ORDER IN a/L AND COMPARISON WITH EXPERIMENTAL DATA

Let us perform calculations with Eq. (34) using the simplest model: the function $W(z)$ is piecewise continuous and takes the values $W_0$ and $W_1$ in the metal and dielectric, respectively, and the electronic spectra of these materials are determined by the equations

$$\epsilon_M(k) = \frac{k^2}{2m}, \quad \epsilon_D(k) = \frac{k^2}{2m} + U, \quad (49)$$

where $U > \epsilon_F$ and $\epsilon_F$ is the Fermi energy in the metal. For a thin layer of dielectric with thickness $d$ inside a metal plate with thickness $L$, the expression for $N(z)$ has the form

$$N(z) = \frac{m}{(2\pi)^2} \int_{q_0}^{k_0} dq \, \frac{q}{k} \, H(k,iq,z)|_{k=\sqrt{k_0^2-q^2}}, \quad (50)$$

where

$$q_0 = \sqrt{k_0^2 - k_F^2}, \quad k_0 = \sqrt{2mU},$$

and the function $H(k,iq,z)$ is defined by Eq. (22) in Ref. 9. Consider two limiting cases corresponding to a narrow-gap semiconductor and a wide-gap dielectric.

(a) $0 < U - \epsilon_F \ll U$.

The result for $U \to \epsilon_F$ and $q_0 \to 0$ coincides with the $q_F \to 0$ limit in Eq. (24) of Ref. 9:

$$\frac{\delta T_c}{T_{c0}} = \frac{1}{\lambda_0 k_F L} \left[\frac{W_1}{W_0} P_1(k_0 d) + P_2(k_0 d)\right], \quad (51)$$

where the functions $P_1(x)$ and $P_2(x)$ are those plotted in Fig. 2 of Ref. 9. This result for finite but small $q_0$ differs from Eq. (51) only in that for $d \gtrsim q_0^{-1} \gg k_0^{-1}$ the algebraic approach to a constant value as $d \to \infty$ described by $P_1(x)$ and $P_2(x)$ is replaced by an exponential behavior. Depending on the ratio $W_1/W_0$, curves of one of the three types shown in Fig. 3a are realized.

(b) $U \gg \epsilon_F$.

In this case $q_0 \approx k_0$, and the limits of integration in Eq. (50) are close. By assuming $q \simeq k_0$ and expanding in terms of $k_F/k_0$, we obtain for $W_1/W_0 \sim k_0^2/k_F^2$

$$\frac{\delta T_c}{T_{c0}} = \frac{1}{\lambda_0 k_F L} \begin{cases} \left(\frac{W_1}{W_0} - \frac{k_0^2}{2k_F^2}\right) k_F d, & k_0 d \ll k_F/k_0, \\ -\frac{2\pi}{3} \frac{k_F^2}{k_0^2} \frac{1}{(k_0 d)^2} \\ + \frac{16}{9} \frac{k_F^5}{k_0^5} \frac{W_1}{W_0} \frac{1}{(k_0 d)^3}, & k_F/k_0 \ll k_0 d \ll 1, \\ \frac{1}{3} \frac{k_F^3}{k_0^3} \left(\frac{W_1}{W_0} \frac{2k_F^2}{3k_0^2} - 1\right) \\ -\frac{8\pi}{3} \frac{k_F^2}{k_0^2} e^{-2k_0 d}, & k_0 d \gg 1. \end{cases}$$
$$(52)$$

Similarly to the previous case, depending on the ratio $W_1/W_0$, the function $T_c(d)$ has one of three typical shapes shown in Fig. 3b. It is remarkable that all three types of curves were recorded by Orlov et al.[14] in Pb–Si, Pb–Ge, and Pb–C sandwiches (Fig. 3c). Since the experiments obviously satisfy $U \sim \epsilon_F$, the experimental curves present an intermediate case between the curves of Figs. 3a and 3b.

Note that for $\bar{\omega}(z) = \text{const}$, when $W(z) = V(z)$, holds the condition $V_1 > V_0$, which is intuitively obvious, is insufficient for increasing $T_c$. A stronger condition is necessary:

$$\frac{V_1}{V_0} > C, \quad C = \begin{cases} 3/2, & U - \epsilon_F \ll \epsilon_F, \\ U/2\epsilon_F, & U \gg \epsilon_F \end{cases}, \quad (53)$$

which is very limiting in the case of a wide-gap dielectric. The point is that for $V(z) = \text{const}$ a spread $\delta$ of the step in the function $N(z)$ defined by Eq. (45) has a negative effect proportional to $\delta$ [see Eq. (34)]. It can be compensated for by the positive effect $\sim \delta(V_1 - V_0)/V_0$ due to the increase in the constant $V$ in the dielectric, and in this case condition (53) with $C > 1$ holds.

## 7. CONCLUSIONS

The issue of the efficiency of the exciton-mediated pairing in layered structures has many aspects, most of which have not been discussed in the paper, namely, whether there are appropriate excitons in the dielectric, whether they penetrate into the metal film to a sufficient depth, whether the excitonic exchange leads to attraction between two electrons, how strong this attraction sufficiently is, etc. The main conclusion of our study is that, even under the most favorable conditions, when the answers to all the above questions are positive (as a result of which, $T_c$ should be high at $L \sim a$), the effect of exciton-mediated pairing would not be detectable at $L \gg a$. Therefore the failure of all the attempts to detect it in sandwiches does not mean that its search in quasi-two-dimensional systems should be abandoned.

1) Modern technologies can produce fairly uniform films with a thickness of several angstroms,[3] but superconductivity in them is suppressed owing to their highly disordered structure that leads to localization effects.[4]

2) Note that this result is not contained in the MacMillan formula, which is, apparently, the main reason why it has not been discovered previously. It can be derived qualitatively from the Eliashberg equations for the homogenous medium if the Eliashberg function is presented in the form $\lambda_{ph}\omega_{ph}\delta(\omega-\omega_{ph})+\lambda_{ex}\omega_{ex}\delta(\omega-\omega_{ex})$ and it is assumed that $\lambda_{ex} \sim a/L$.

3) Notice that $\lambda$ is defined in terms of the BCS theory; the values $\lambda \sim 1$ are obtained using MacMillan-type formulas, in which $1/\lambda_{BCS}$ is replaced by a combination of the form $(1+\lambda)/(\lambda-\mu^*)$, where $\mu^*$ is the Coulomb pseudopotential.[2]

4) Application of this technique directly to the Matsubara representation is notably easier than with a preliminary analytic continuation (Ref. 2, Ch. 4) and yields identical results.

[1] V. A. Ginzburg, Zh. Éksp. Teor. Fiz. **47**, 2318 (1964) [Sov. Phys. JETP **20**, 1549 (1965)]; Phys. Lett. **13**, 101 (1964).
[2] *High-Temperature Superconductivity*, ed. by V. L. Ginzburg and D. A. Kirzhnits, Consultants Bureau, New York, 1982.
[3] S. M. Durkin, J. E. Cunningham, M. E. Mochel, and C. P. Flynn, J. Phys. F **11**, L223 (1981).
[4] B. N. Belevtsev, Usp. Fiz. Nauk **160**, 65 (1990) [Sov. Phys. Uspekhi **33**, 36 (1990)].
[5] V. L. Ginzburg and D. A. Kirzhnits, Zh. Éksp. Teor. Fiz. **46**, 397 (1964)[Sov. Phys. JETP **19**, 219 (1964)].
[6] I. M. Suslov, Zh. Éksp. Teor. Fiz. **95**, 949 (1989) [Sov. Phys. JETP **68**, 546 (1989)].
[7] I. M. Suslov, Sverkhprovodimist': Fizika, Khimiya, Tekhnologiya **4**, 2093 (1991).
[8] Yu. A. Krotov and I. M. Suslov, Zh. Éksp. Teor. Fiz. **102**, 670 (1992); **103**, 1394 (1993); Physica C **213**, 421 (1993); **245**, 252 (1995).
[9] Yu. A. Krotov and I. M. Suslov, Zh. Éksp. Teor. Fiz. **107**, 512 (1995). [Sov. Phys. JETP **80**, 275 (1995)].
[10] A. A. Abrikosov, L. P. Gor'kov, and I. E. Dzyaloshinskii, *Methods of Quantum Field Theory in Statistical Physics*, Prentice-Hall, Englewood Cliffs, NJ, 1963.
[11] P. G. De Gennes, Rev. Mod. Phys. **36**, 225 (1964).
[12] J. Ziman, *Electrons and Phonons*, Oxford University Press (1960).
[13] A. A. Maradudin, E. W. Montroll, G. H. Weiss, and I. P. Ipatova, *Theory of Lattice Dynamics in the Harmonic Approximation*, Academic Press, New York–London (1971).
[14] A. F. Orlov, A. K. Milai, and V. P. Dmitriev, Fiz. Tverd. Tela **18**, 1470 (1976) [Sov. Phys. Solid State **18**, 854 (1976)].

Translation was provided by the Russian Editorial office.